AUTHORS:

**Daniela Vorkel**

Myers lab, Center for Systems Biology

Max Planck Institute for Molecular Cell Biology and Genetics, Dresden

**Robert Haase**

Myers lab, Center for Systems Biology

Max Planck Institute for Molecular Cell Biology and Genetics, Dresden

# GPU-accelerating ImageJ Macro image processing workflows using CLIJ

**August 2020**

# Contents

## 1.1 What you will learn in this chapter

This chapter introduces GPU-accelerated image processing in ImageJ/FIJI. The reader is expected to have some pre-existing knowledge of ImageJ Macro programming. Core concepts such as variables, *for*-loops, and functions are essential. The chapter provides basic guidelines for improved performance in typical image processing workflows. We present in a step-by-step tutorial how to translate a pre-existing ImageJ macro into a GPU-accelerated macro.

## 1.2 Introduction

Modern life science increasingly relies on microscopic imaging followed by quantitative bio-image analysis (BIA). Nowadays, image data scientists join forces with artificial intelligence researchers, incorporating more and more machine learning algorithms into BIA workflows. Despite general machine learning and convolutional neural networks are not new approaches to image processing, they are of increasing importance for life science. As their application is now at hand due to the rise of advanced computing hardware, namely graphics processing units (GPUs), poses the question if GPUs can also be exploited for classic image processing in ImageJ (Schneider et al., 2012) and Fiji (Schindelin et al., 2012). As an alternative to established acceleration techniques, such as the batch mode, we explored how GPUs can be exploited to accelerate classic image processing. Our approach, called CLIJ (Haase et al., 2020), enables biologists and bio-image analysts to speed up time-consuming analysis tasks by adding OpenCL-support (Khronos-Group, 2020) to ImageJ. We present a guide for transforming state-of-the-art image processing workflows into GPU-accelerated workflows using the ImageJ Macro language. Our suggested approach neither requires a profound expertise in high performance computing, nor to learn a new programming language such as OpenCL.

To demonstrate the procedure, we translate a formerly published BIA workflow for examining signal intensity changes at the nuclear envelope, caused by cytoplasmic redistribution of a fluorescent protein (Miura, 2020). There-fore, we introduce ways to discover CLIJ commands as counterparts of classic ImageJ methods. These commands are then assembled to refactor the pre-existing workflow. In terms of image processing, refactoring means to restructure an existing macro without changing measurement results, and rather to improve processing speed. Accordingly, we show how to measure workflow performance. We also give an insight into quality assurance methods, which help to ensure good scientific practice when modernizing BIA workflows and refactoring code.

## 1.3 The data set

**Imaging data**

Looking at cells, membranes create functional compartments and maintain diverse content and activities. Fluorescent labeling techniques allow to study certain structures and cell components, in particular to trace dynamic processes over time, such as changes in intensity and spatial distribution of fluorescent signals. The method of live-imaging, taken as long-term time lapses, became important to study dynamic biological processes. As a representative data set for this domain, we process a two-channel time-lapse showing a Hela Cell with increasing signal intensity in one channel (Boni et al., 2015). The data set has a pixel size of 0.165 um per pixel and a frame rate of 400 seconds per frame. The nuclei-channel (C1), excited with 561 nm wavelength, consists of H2B-mCherry signals within the nucleus. The protein-channel (C2), excited with 488 nm wavelength, represents the distribution of the cytoplasmic Lamin B protein, which allocates from the endoplasmic reticulum to the inner nuclear membrane (Lamin B receptor signal). Four example time points of the data set are shown in Fig. 1.1.

**The predefined processing workflow**

To measure changing intensities along the nuclear envelope, it is required to define a corresponding region of interest (ROI) within the image: First, the image is segmented into nucleus and background. Second, a region

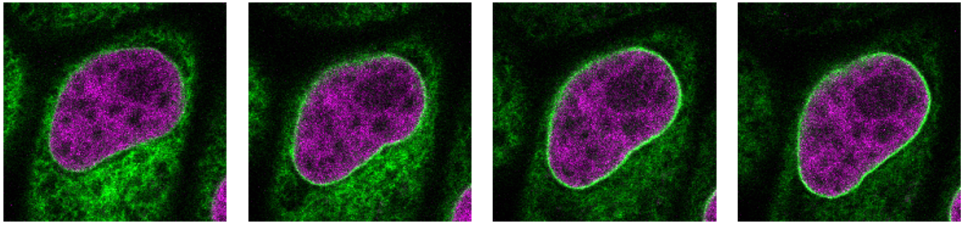

Figure 1.1: Time points 1, 5, 10 and 15 of the example data set showing signal increase in the nuclear envelope. Courtesy: Andrea Boni, EMBL Heidelberg/Viventis.

surrounding the nucleus is derived.

Starting point for the workflow translation is the *code_final.ijm* macro file published by Miura (2020) [1]. For reader's convenience, we added explaining comments for each section:

```
1  // determine current data set and split channels
2  orgName = getTitle();
3  run("Split Channels");
4  c1name = "C1-" + orgName;
5  c2name = "C2-" + orgName;
6
7  // invoke segmentation of a band around the nucleus
8  selectWindow(c1name);
9  nucorgID = getImageID();
10 nucrimID = nucseg( nucorgID );
11
12 // go through all time points and measure intensity in the
       band
13 selectWindow(c2name);
14 c2id = getImageID();
15 opt = "area mean centroid perimeter shape integrated
       display redirect=None decimal=3";
16 run("Set Measurements...", opt);
17 for (i =0; i < nSlices; i++){
18     selectImage( nucrimID );
19     setSlice( i + 1 );
20     run("Create Selection");
21     run("Make Inverse");
22     selectImage( c2id );
```

---

[1]https://github.com/miura/NucleusRimIntensityMeasurementsV2/
blob/master/code/code_final.ijm

```
23      setSlice( i + 1 );
24      run("Restore Selection");
25      run("Measure");
26 }
27
28 // detailed segmentation of the band around the nucleus
29 function nucseg( orgID ){
30      selectImage( orgId );
31      run("Gaussian Blur...", "sigma=1.50 stack");
32      setAutoThreshold("Otsu dark");
33      setOption("BlackBackground", true);
34      run("Convert to Mask", "method=Otsu background=Dark
            calculate black");
35      run("Analyze Particles...", "size=800-Infinity pixel
            circularity=0.00-1.00 show=Masks display exclude
            clear include stack");
36      dilateID = getImageID();
37      run("Invert LUT");
38      options =  "title = dup.tif duplicate range=1-" +
            nSlices;
39      run("Duplicate...", options);
40      erodeID = getImageID();
41      selectImage(dilateID);
42      run("Options...", "iterations=2 count=1 black edm=
            Overwrite do=Nothing");
43      run("Dilate", "stack");
44      selectImage(erodeID);
45      run("Erode", "stack");
46      imageCalculator("Difference create stack", dilateID,
            erodeID);
47      resultID = getImageID();
48      selectImage(dilateID);
49      close();
50      selectImage(erodeID);
51      close();
52      selectImage(orgID);
53      close();
54      run("Clear Results");
55      return resultID;
```

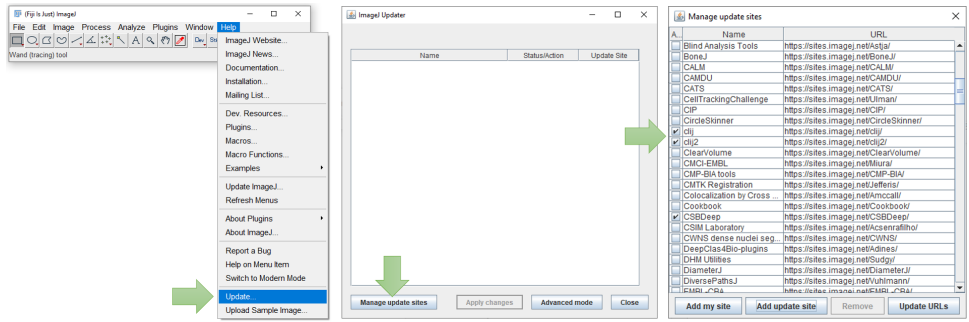

Figure 1.2: Installation of CLIJ: In Fiji's updater, which can be found in the menu *Help > Update...*, click on *Manage Update Sites*, activate the checkboxes next to clij and clij2. After updating and restarting Fiji, CLIJ is installed.

```
56 }
```

code/code_original.ijm

## 1.4 Tools: CLIJ

Available as optional plugin, CLIJ brings GPU-accelerated image processing routines to Fiji. Installation of CLIJ is done by using the Fiji updater, which can be found in the menu *Help > Update*, and by activating the update sites of clij and clij2, as shown in Fig. 1.2. Dependent on GPU vendor and operating system, further installation of GPU drivers might be necessary. In some cases, default drivers delivered by automated operating system updates are not sufficient.

After installing CLIJ, it is recommended to execute a CLIJ macro to test for successful installation. We can also use this opportunity to get a first clue about a potential speedup of a CLIJ method compared to its ImageJ counterpart. The following example macro processes an image using both methods, and writes the processing time into the log window, as shown in Fig. 1.3.

```
1 // load example dataset
2 run("T1 Head (2.4M, 16-bits)");
3
4 // initialize GPU
5 run("CLIJ2 Macro Extensions", "cl_device=");
```

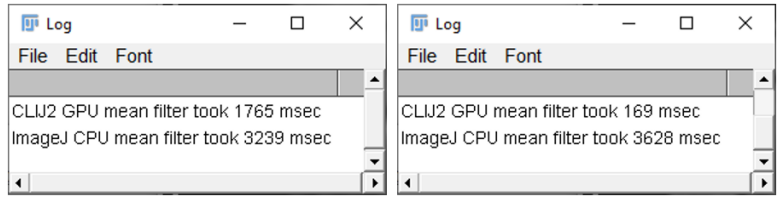

Figure 1.3: Output of the first example macro, which reports processing time of a CLIJ operation (first line), and of the classic ImageJ operation (second line). When executing a second time (right), the GPU typically becomes faster due to the so called warm-up effect.

```
 6 Ext.CLIJ2_clear();
 7
 8 // apply a mean filter on the GPU
 9 time = getTime();
10 input = getTitle();
11 Ext.CLIJ2_push(input);
12 Ext.CLIJ2_mean3DBox(input, result, 3, 3, 3);
13 Ext.CLIJ2_pull(result);
14 Ext.CLIJ2_clear();
15 print("CLIJ2 GPU mean filter took " + (getTime() - time) +
       " msec");
16
17 // apply the corresponding operation of classic ImageJ
18 time = getTime();
19 run("Mean 3D...", "x=3 y=3 z=3");
20 print("ImageJ CPU mean filter took " + (getTime() - time) +
        " msec");
```
<div align="center">code/first_example.ijm</div>

**Basics of GPU-accelerated image processing with CLIJ**

Every ImageJ macro, which uses CLIJ functionality, needs to contain some additional code sections. For example, this is how the GPU is initialized:

```
run("CLIJ2 Macro Extensions", "cl_device=");
Ext.CLIJ2_clear();
```

In the first line, the parameter *cl_device* can stay blank, imposing that CLIJ will select automatically an OpenCL device, namely the GPU. One can specify the name of the GPU in brackets, for example *nVendor Awesome Intelligent*. If only a part of the name is specified, such as *nVendor* or *some*, CLIJ will select a GPU which contains that part of the name. One can explore available GPU devices by using the menu *Plugins > ImageJ on GPU (CLIJ2) > Macro tools > List available GPU devices*. The second line, in the example shown above, cleans up GPU memory. This command is typically called by the end of a macro. It is not mandatory to write it at the beginning. However, it is recommended while elaborating a new ImageJ macro. A macro under development unintentionally stops every now and then with error messages. Hence, a macro is not executed until the very end, where GPU memory typically gets cleaned up. Thus, it is recommended to write this line initially, to start at a predefined empty state.

Another typical step in CLIJ macros is to push image data to the GPU memory:

```
input = getTitle();
Ext.CLIJ2_push(input);
```

We first retrieve the name of the current image by using ImageJs built-in *getTitle()*-command, save it into the variable input. Afterwards, the *input* image is stored in GPU memory using CLIJs push method.

This image can then be processed, for example using a mean filter:

```
Ext.CLIJ2_mean3DBox(input, result, 3, 3, 3);
```

CLIJ's mean filter, applied to a 3D image, takes a cuboidal neighborhood into account, as specified by the word *Box*. It has five parameters: the *input* image name, the *result* image name given by variables, and three half-axis lengths describing the size of the box. If the variable for the result is not set, it will be set to an automatically generated image name.

Finally, the *result*-image gets pulled back from GPU memory and will be displayed on the screen.

```
Ext.CLIJ2_pull(result);
Ext.CLIJ2_clear();
```

Hence, result images are not shown on the screen until the *pull()*-command is explicitly called. Thus, the computer screen is not flooded with numerous image windows, helping the workflow developer to stay organised. Furthermore, memory gets cleaned up by the *clear()*-command, as explained above.

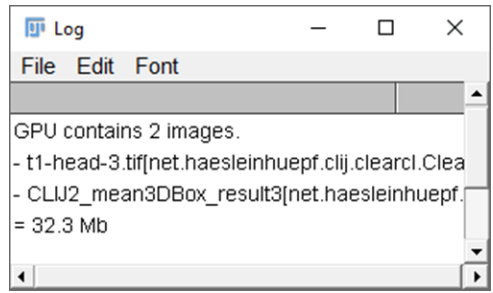

Figure 1.4: List of images currently stored in GPU memory: In this case, there exists an image called *t1-head-3.tif* which corresponds to the dataset we loaded initially. Furthermore, there is another image, called *CLIJ2_mean3DBox_result3*, obviously containing the result of the mean filter operation.

While developing advanced CLIJ workflows, it might be necessary to take a look into GPU memory to figure out, which images are stored at a particular moment. Therefore, we can add another command just before the final *clear()*-command, which will list images in GPU memory in the log windows, as shown in Fig. 1.4:

```
Ext.CLIJ2_reportMemory();
```

As intermediate summary, CLIJ commands in ImageJ macro typically appear like this:

```
Ext.CLIJ2_operation(parameters);
```

All CLIJ methods start with the prefix *Ext.*, a convention by classical ImageJ, indicating that we are calling a macro extension optionally installed to ImageJ. Next, it reads *CLIJ_*, *CLIJ2_* or *CLIJx_* followed by the specific method, and in brackets the parameters passed over to this method. Parameters are typically given in the order: input images, output images, other parameters.

The *CLIJ* identifier was introduced to classify methods originally published as CLIJ toolbox (Haase et al., 2020). It is now deprecated since the official stable release of *CLIJ2*, which is the extended edition with more operations. Furthermore, there is *CLIJx*, the volatile experimental sibling, which is constantly evolving as developers work on it. Thus, CLIJx methods should be used with care as the *X* stands for e*X*perimental. Whenever possible, the latest stable release should be used. As soon as a new stable release is out, the former one will be deprecated. The deprecated release will be kept available for at least one year. To allow a convenient transition between major releases, the CLIJ developers strive for backwards-compatibility between releases.

**Where CLIJ is conceptually different and why**

When designing the CLIJ application programming interface (API), special emphasis was put on a couple of aspects to standardize and simplify image processing.

- Results of CLIJ operations are per default not shown on screen. One needs to pull the image data from the GPU memory to display it in an ImageJ window. In order to achieve optimal performance, it is recommended to execute as many processing steps as possible between push and pull commands. Furthermore, only the final result image should be pulled. Pushing and pulling take time. This time investment can be gained back by calling operations, which are altogether faster than the classic ImageJ operations.

- CLIJ operations always need explicit specifications of input and output images. The currently selected window in ImageJ does not play a role when calling a CLIJ command. Moreover, no command in CLIJ changes the input image. All commands read pixels from input images and write new pixels into output images:
  ```
  Ext.CLIJ2_excludeLabelsOnEdges(labels,
      labels_without_touching_edges);
  ```

- CLIJ operations do not take physical units into account. For example, all radius and sigma parameters are provided in pixel units:
  ```
  sigma = 1.5;
  Ext.CLIJ2_gaussianBlur2D(orgID, blurred, sigma,
      sigma);
  ```

- If a CLIJ method's name contains the terms "2D" or "3D", it respectively processes images with two or three dimensions. If the name of the method is without such a term, the method processes images of both kind.

- Images and image stacks in CLIJ are the granular units of data. On the one hand, accessing individual pixels cannot be done efficiently on GPUs. Hence, it is recommended to use classic ImageJ functions for that. On the other hand, time lapse data need to be split into image stacks and processed time point by time point in a for-loop.

- CLIJ methods are granular operations. They apply a single defined procedure to a whole image. An operation never involves any additional operation depending on given parameters. Furthermore, independent from any ImageJ configuration, CLIJ methods produce the same output given the same input. Hence, CLIJ methods are not influenced by side-effects. This side-effect free and granular operation concept leads to improved readability and maintenance of image processing workflows.

**Hardware suitable for CLIJ**

When using CLIJ for best possible performance, it is recommended to use recent GPUs. Technically, CLIJ is compatible with GPU-devices supporting the OpenCL 1.2 standard ([Khronos-Group, 2020](#)), which was established in 2011. While OpenCL works on GPUs up to 9 years old, GPU devices older than 5 years may be unable to offer a processing performance faster than recent CPUs. Thus, when striving for high performance, recent devices should be utilized. When considering new hardware, image processing specific aspects should be taken into account:

- **Memory size**: State-of-the-art imaging techniques produce granular 2D and 3D image data up several gigabytes. Dependent on the desired use case, it may make sense to utilize GPUs with increased memory. Convenient workflow development is possible, if a processed image fits about 4-6 times into GPU memory. Hence, if working with images of 1-2 GB in size, a GPU with at least 8 GB of GDDR6 RAM memory should be used.

- **Memory Bandwidth**: Image processing is memory-bound, meaning that all operations have in common that pixels are read from memory and written to memory. Reading and writing is the major bottleneck, and thus, GPUs with fast memory access and with high memory bandwidth should be preferred. Typically, GDDR6-based GPUs have memory bandwidths larger than 400 GB/s. GDDR5-based GPUs often offer less than 100 GB/s. So, GDDR6-based GPUs may compute image processing results about 4 times faster.

- **Integrated GPUs**: For processing of big images, a large amount of memory might be needed. As of writing this, GDDR6-based GPUs with 8 GB of memory are available in price ranges between 300 and 500 EUR. GPUs with more than 20 GB of memory cost about ten fold. Despite drawbacks in processing speed, it also might make sense to use integrated GPUs with access to huge amounts of DDR4-memory.

## 1.5 The Workflow

### 1.5.1 Macro translation

The CLIJ Fiji plugin and its individual CLIJ operations were programmed in a way that ImageJ users find well-known concepts when translating workflows, and can use CLIJ operations as if they were ImageJ operations. There are some differences, aiming at improved user experience, we would like to highlight in this section.

**The Macro Recorder**

The ImageJ macro recorder is one of the most powerful tools in ImageJ. While the user calls specific menus to process images, it automatically records code. The recorder is launched from the menu *Plugins -> Macros -> Record....* The user can call any CLIJ operation from the menu. For example, the first step in the nucleus segmentation workflow is applying a Gaussian blur to a 2D image. This operation can be found in the menu *Plugins > ImageJ on GPU (CLIJ2) > Filter > Gaussian blur 2D on GPU*. When executing this command, the macro recorder will record this code:

```
run("CLIJ2 Macro Extensions", "cl_device=[Intel(R) UHD
   Graphics 620]");

// gaussian blur
image1 = "NPCsingleNucleus.tif";
Ext.CLIJ2_push(image1);
image2 = "gaussian_blur-1901920444";
sigma_x = 2.0;
sigma_y = 2.0;
Ext.CLIJ2_gaussianBlur2D(image1, image2, sigma_x, sigma_y);
Ext.CLIJ2_pull(image2);
```

All recorded CLIJ-commands follow the same scheme: The first line initializes the GPU, and explicitly specifies the used OpenCL device while executing a operation. The workflow developer can remove this explicit specification as introduced in section 1.4. Afterwards, the parameters of the command are listed and specified. Input images, such as *image1* in the example above, are pushed to the GPU to have them available in its memory. Names are assigned to output image variables, such as *image2*. These names are automatically generated and supplemented with a unique number in the name. The developer is welcome to edit these names to improve code readability. Afterwards, the operation *GaussianBlur2D* is executed on the GPU. Finally, the resulting image is pulled back from GPU memory to be visualized on the screen as an image window.

**Fijis search bar**

As ImageJ and CLIJ come with many commands and in huge menu structures, one may not know in which menu specific commands are listed. To search for commands in Fiji, the Fiji search bar is a convenient tool as shown in Fig. 1.5 a. For example, the next step in our workflow is segmenting the blurred image using an histogram-based thresholding algorithm (Otsu, 1979). When entering *Otsu* in the search field, related commands will be listed in the search result. Hitting the *Enter* key or clicking the *Run* button will execute the command as if it was called from the menu. Hence, also identical code will be recorded in the macro recorder.

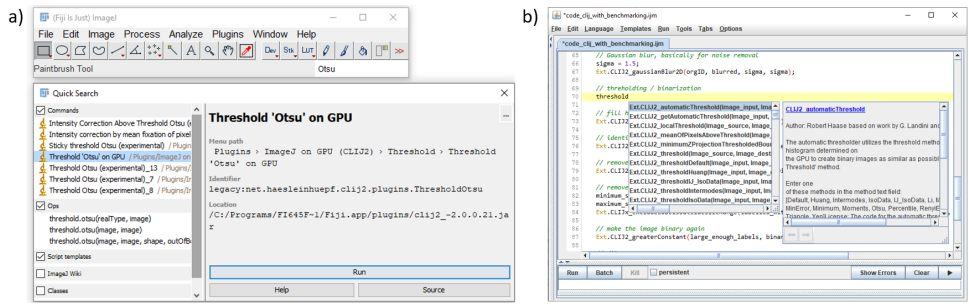

Figure 1.5: a) While recording macros, the Fiji search bar helps to discover CLIJ commands in the menu. b)Auto-Completion in Fijis script editor supports the workflow developer in discovering commands and offers their documentation.

**The script editor and the auto-complete function**

In the Macro Recorder window, there is a *Create*-button which opens the Script Editor. In general, it is recommended to record a gross workflow. To extend code, to configure parameters, and to refine execution order, one should switch to the Script Editor. The script editor exposes a third way for exploring available commands: The auto-complete function, shown in Fig. 1.5 b. Typing *threshold* will open two windows: A list of commands which contain the searched word. The position of the searched word within the command does not matter. Thus, entering *threshold* or *otsu* will both lead to the command *thresholdOtsu*. Furthermore, a second window will show the documentation of the respectively selected command. By hitting the *Enter* key the selected command is auto-completed in the code, for example like this:

```
Ext.CLIJ2_thresholdOtsu(Image_input, Image_destination);
```

The developer can then replace the written parameters *Image_input* and *Image_destination* with custom variables.

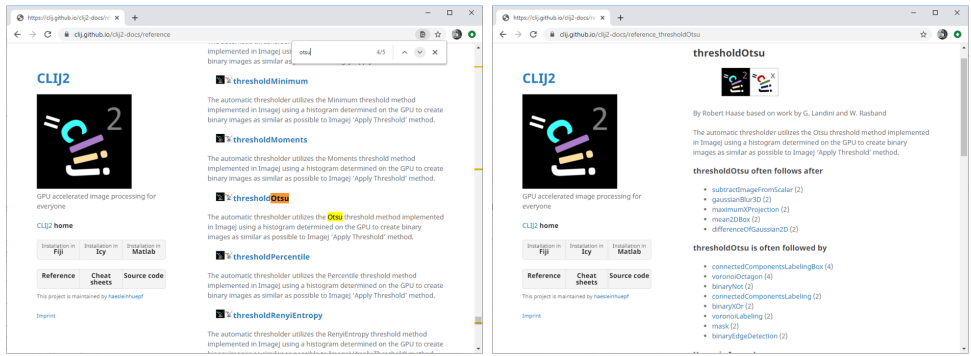

Figure 1.6: The online API reference can be explored using the search function of the internet browser, e.g. for algorithms containing *Otsu*(left). The documentation of specific commands contains a list of typical predecessors and successors (right). For example, thresholding is typically followed by connected components labelling, the core algorithm behind ImageJs Particle Analyzer.

**The CLIJ website and API reference**

Furthermore, the documentation window of the auto-complete function is connected to the API reference section of the CLIJ website[2], as shown in Fig. 1.6. The website provides a knowledge base, holding a complete list of operations and typical workflows connecting operations with each other. For example, this becomes crucial when searching for the CLIJ analog of the ImageJs Particle Analyser as there is no such operation in CLIJ. The website lists typical operations following Otsu-thresholding, for example connected components labelling, the core-algorithm behind ImageJs Particle Analyzer.

**Exercise 1**

Open the Macro Recorder and the example image *NPCsingleNucleus.tif*. Type *Otsu* into the Fiji search bar. Select the CLIJ2 method on GPU and run the threshold using the button *Run*. Read in the online documentation which commands are typically applied before Otsu thresholding. Which of those commands can be used to improve the segmentation result?

---

[2]https://clij.github.io/reference

### 1.5.2 The new workflow routine

While reconstructing the workflow, this tutorial follows the routines of the classic macro, and restructures the execution order of commands to prevent minor issues with pre-processing before thresholding. The processed dataset is a four-dimensional data set, consisting of spatial dimensions X and Y, channels and frames. When segmenting the nuclear envelope in the original workflow, the first operation applied to the data set is a *Gaussian blur*:

```
run("Gaussian Blur...", "sigma=1.50 stack");
```

The *stack* parameter suggests that this operation is applied to all time points in both channels, potentially harming later intensity measurements. However, for segmentation of the nuclear envelope in a single time point image, this is not necessary. As introduced in section 1.4, such kind of data is not of granular nature and has to be cut into 2D images before applying CLIJ operations. We can use the method *pushCurrentSlice* to push a single 2D image to the GPU memory. Then, a 2D segmentation can be generated, using a workflow similar to the originally proposed workflow. Finally, we pull the segmentation back as ROI and perform statistical measurements using classic ImageJ. Thus, the content of the *for*-loop in the original program needs to be reorganized:

```
for (i = 0; i < frames; i ++) {

    // navigate to a given time point in our stack
    Stack.setFrame(i + 1);

    // select the channel showing nuclei
    Stack.setChannel(nuclei_channel);

    // get a single-channel slice
    Ext.CLIJ2_pushCurrentSlice(orgName);

    // segment the nuclear envelope
    nucrimID = nucseg( orgName );

    // select the channel showing nuclear envelope signal
    Stack.setChannel(channel_to_measure);
```

```
    // pull segmented binary image as ROI from GPU
    Ext.CLIJ2_pullAsROI(nucrimID);

    // analyse it
    run("Measure");

    // remove selection
    run("Select None");
}
```

The function *nucseg* takes an image from the nucleus channel and segments its nuclear envelope. Table 1.1 shows translations from original ImageJ macro functions to CLIJ operations.

While the translation of commands for thresholding is straightforward, other translations need to be explained in more detail, for example the *Analyze Particles* command:

```
run("Analyze Particles...", "size=800-Infinity pixel
    circularity=0.00-1.00 show=Masks display exclude clear
    include stack");
```

The advanced ImageJ macro programmer knows that this line does post-processing of the thresholded binary image, and executes in fact five operations: 1) It identifies individual objects in the binary image, also known as connected components labeling. 2) It removes objects smaller than 800 pixels (*size=800-Infinity pixel*). 3) It removes objects touching the image edges (*exclude*). 4) It fills black holes in white areas (*include*), and 5) it finally converts the image again into a binary image (*show=Masks*). The remaining parameters of the command, *circularity=0.00-1.00*, *display* and *clear*, are in fact not relevant for this processing step, or specify that the operations should be applied to the whole stack (*stack*) slice-by-slice. Thus, the parameters specify commands, which should be executed, but they are not given in execution order. As explained in section 1.4, CLIJ operations are granular. When working with CLIJ, each of the five operations above must be executed explicitly and in the right order. This leads to longer code, but is easier to read and to maintain.

```
// Fill black holes in white objects
```

| ImageJ Macro | ImageJ + CLIJ Macro |
|---|---|
| **Gaussian Blur** | |
| `run("Gaussian Blur...", "sigma=1.50 stack" );` | `sigma = 1.5;`<br>`Ext.CLIJ2_gaussianBlur2D(orgID, blurred, sigma, sigma);` |
| **Thresholding (Otsu, 1979) and analyze particles to eliminate small objects** | |
| `setAutoThreshold("Otsu dark");`<br>`setOption("BlackBackground", true);`<br>`run("Convert to Mask", "method=Otsu background=Dark calculate black");`<br>`run("Analyze Particles...", "size=800-Infinity pixel circularity=0.00-1.00 show=Masks display exclude clear include stack");` | `Ext.CLIJ2_thresholdOtsu(blurred, thresholded);`<br>`Ext.CLIJ2_binaryFillHoles(thresholded, holes_filled);`<br>`Ext.CLIJ2_connectedComponentsLabelingBox(holes_filled, labels);`<br>`Ext.CLIJ2_excludeLabelsOnEdges(labelled, labels_wo_edges);`<br>`Ext.CLIJx_excludeLabelsOutsideSizeRange(labels_wo_edges, large_enough_labels, 800, 1000000);`<br>`Ext.CLIJ2_greaterConstant(large_enough_labels, binary_mask, 0);` |
| **Dilation** | |
| `run("Options...", "iterations=2 count=1 black edm=Overwrite do=Nothing");`<br>`run("Dilate", "stack");` | `radius = 2;`<br>`Ext.CLIJ2_maximum2DBox(binary_mask, dilateID, radius, radius);` |
| **Erosion** | |
| `run("Erode", "stack");` | `Ext.CLIJ2_minimum2DBox(binary_mask, erodeID, radius, radius);` |
| **Image subtraction** | |
| `imageCalculator("Difference create stack", dilateID, erodeID);` | `Ext.CLIJ2_subtractImages(dilateID, erodeID, resultID);` |
| **ROI generation** | |
| `selectImage( nucrimID );`<br>`run("Create Selection");`<br>`selectImage( c2id );`<br>`run("Restore Selection");` | `Ext.CLIJ2_pullAsROI(nucrimID);` |

Table 1.1: ImageJ macro to CLIJ macro translations in the context of the example workflow.

```
Ext.CLIJ2_binaryFillHoles(thresholded, holes_filled);

// Identify individual objects
Ext.CLIJ2_connectedComponentsLabelingBox(holes_filled,
   labels);

// Remove objects which touch the image edge
Ext.CLIJ2_excludeLabelsOnEdges(labels, labels_wo_edges);

// Exclude objects smaller than 800 pixels
minimum_size = 800;
maximum_size = 1000000; // large number
Ext.CLIJx_excludeLabelsOutsideSizeRange(labels_wo_edges,
   large_labels, minimum_size, maximum_size);

// generate a new binary image
Ext.CLIJ2_greaterConstant(large_labels, binary_mask, 0);
```

The whole translated workflow looks then like this:

```
1  // configure channels
2  nuclei_channel = 1;
3  protein_channel = 2;
4
5  // Initialize GPU
6  run("CLIJ2 Macro Extensions", "cl_device=");
7  Ext.CLIJ2_clear();
8
9  // determine current image
10 orgName = getTitle();
11
12 // configure measurements (on CPU)
13 opt = "area mean centroid perimeter shape integrated
       display redirect=None decimal=3";
14 run("Set Measurements...", opt);
15
16 getDimensions(width, height, channels, slices, frames);
17 for (i = 0; i < frames; i ++) {
18    // select channel and frame to analyze
19    Stack.setChannel(nuclei_channel);
20    Stack.setFrame(i + 1);
```

```
21
22      // get a single-channel slice
23      Ext.CLIJ2_pushCurrentSlice(orgName);
24
25      // segment the nuclear envelope
26      nucrimID = nucseg( orgName );
27
28      // select the channel showing nuclear envelope signal
29      Stack.setChannel(protein_channel);
30
31      // pull segmented binary image as ROI from GPU
32      Ext.CLIJ2_pullAsROI(nucrimID);
33
34      // analyse it
35      run("Measure");
36
37      // reset selection
38      run("Select None");
39 }
40
41 // This function segments the nuclear envelope in the
      nuclei-channel
42 function nucseg( orgID ){
43      // Gaussian blur, basically for noise removal
44      sigma = 1.5;
45      Ext.CLIJ2_gaussianBlur2D(orgID, blurred, sigma, sigma);
46
47      // thresholding / binarization
48      Ext.CLIJ2_thresholdOtsu(blurred, thresholded);
49
50      // fill holes
51      Ext.CLIJ2_binaryFillHoles(thresholded, holes_filled);
52
53      // identify individual objects
54      Ext.CLIJ2_connectedComponentsLabelingBox(holes_filled,
         labels);
55
56      // remove objects which touch image border
57      Ext.CLIJ2_excludeLabelsOnEdges(labels, labels_wo_edges)
         ;
58
```

```
59      // remove objects out of a given size range
60      minimum_size = 800;
61      maximum_size = 1000000;
62      Ext.CLIJx_excludeLabelsOutsideSizeRange(labels_wo_edges
            , large_labels, minimum_size, maximum_size);
63
64      // make the image binary again
65      Ext.CLIJ2_greaterConstant(large_labels, binary_mask, 0)
            ;
66
67      // dilate
68      radius = 2;
69      Ext.CLIJ2_maximum2DBox(binary_mask, dilateID, radius,
            radius);
70
71      // erode
72      Ext.CLIJ2_minimum2DBox(binary_mask, erodeID, radius,
            radius);
73
74      // subtract eroded from dilated image to get a band
            corresponding to nuclear envelope
75      Ext.CLIJ2_subtractImages(dilateID, erodeID, resultID);
76
77      // return result
78      return resultID;
79 }
```

<div align="center">code/code_clij_final.ijm</div>

**Further optimization**

So far, we translated a pre-existing segmentation workflow without changing processing steps, and with the goal of replicating results. If processing speed plays an important role, it is possible to further optimize the workflow, accepting that results may be slightly different. Therefore, it is necessary to identify code sections which have a high potential for further optimization. To trace down the time consumption of code sections, we now introduce three more CLIJ commands:

```
Ext.CLIJ2_startTimeTracing();
```

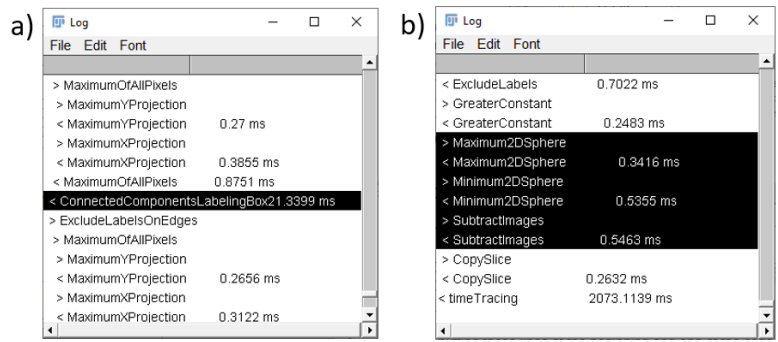

Figure 1.7: An example of printed time traces reveals that a) connected components labeling takes about 21 ms per slice and binary erosion, dilation and subtraction of images takes about 1.3 ms per slice.

```
// here comes the workflow we want to analyze

Ext.CLIJ2_stopTimeTracing();
Ext.CLIJ2_getTimeTracing(time_traces);
print(time_traces);
```

By including these lines at the beginning and the end of a macro, we can trace elapsed time during command executions in the log window, as shown in Fig. 1.7. In that way, one can identify parts of the code where most of the time is spent. In case of the implemented workflow, connected components labelling reveals as bottleneck. In order to exclude objects smaller than 800 pixels from segmentation, we need to call connected components labelling. By skipping this step and accepting a lower quality of segmentation, we could have a faster processing. This leads to a shorter workflow:

```
function nucseg( orgID ){
    // blur the image to get a smooth outline
    sigma = 1.5;
    Ext.CLIJ2_gaussianBlur2D(orgID, blurred, sigma, sigma);

    // threshold it
    Ext.CLIJ2_thresholdOtsu(blurred, thresholded);

    // fill holes in the binary image
    Ext.CLIJ2_binaryFillHoles(thresholded, binary_mask);
```

```
    // dilate the binary image
    radius = 2;
    Ext.CLIJ2_maximum2DBox(binary_mask, dilateID, radius,
        radius);

    // erode the binary image
    Ext.CLIJ2_minimum2DBox(binary_mask, erodeID, radius,
        radius);

    // subtract the eroded from the dilated image
    Ext.CLIJ2_subtractImages(dilateID, erodeID, resultID);
    return resultID;
}
```

Analogously, an optimization is also thinkable for the classic workflow. When executing the optimized version of the two workflows, we retrieve different measurements, which will be discussed in the following section.

**Exercise 2**

Start the ImageJ Macro Recorder, open an ImageJ example image by clicking the menu *File > Open Samples > T1 Head (2.4M, 16 bit)* and apply the *Top Hat* filter to it. In the recorded ImageJ macro, activate time tracing before calling the *Top Hat* filter to study what is actually executed when running the *Top Hat* operation and how long it takes. What does the *Top Hat* operation do?

### 1.5.3 Good scientific practice in method comparison studies

When refactoring scientific image analysis workflows, good scientific practice includes quality assurance to check if a new version of a workflow produces identical results with given tolerance. In software engineering, the procedure is known as regression testing. Translating workflows for the use of GPUs instead of CPUs, is one example to do so. In a wider context, other examples are switching major software versions, operating systems, CPU or GPU hardware, or computational environments, such as ImageJ and Python.

Starting from a given data set, we can execute a reference script to generate reference results. Running a refactored script, or executing a script under different conditions will deliver new results. To compare these results to the reference, we use different strategies, ordered from the simplest to the most elaborated approach: 1) comparison of mean values and standard deviation, 2) correlation analysis, 3) equivalence testing, and 4) Bland-Altman analysis. For demonstration purpose, we apply these strategies to our four workflows:

- W-IJ: Original ImageJ workflow
- W-CLIJ: Translated CLIJ workflow
- W-OPT-IJ: Optimized ImageJ workflow
- W-OPT-CLIJ: Optimized CLIJ workflow

In addition, we execute the CLIJ macros on four computers with different CPU/GPU specifications:

- Intel i5-8265U CPU/ Intel UHD 620 integrated GPU
- Intel i7-8750H CPU/ NVidia Geforce 2080 Ti RTX external GPU
- AMD Ryzen 4700U CPU/ AMD Vega 7 integrated GPU
- Intel i7-7920HQ CPU/ AMD Radeon Pro 560 dedicated GPU

**Comparison of mean values and standard deviation**

An initial and straightforward strategy is to compare mean and standard deviation of workflows. If the difference between then mean measurements exceeds a given tolerance, the new workflow cannot be utilized to study the phenomenon as done by the original workflow. However, if means are equal or very similar, this does not allow to conclude that the methods are exchangeable. Close mean and standard deviation values are necessary, but not sufficient to prove method similarity. Results of the method comparison, using mean and standard deviation, are shown in Table 1.2.

| Workflow | Intel CPU Intel iGPU | Intel CPU NVidia eGPU | AMD CPU AMD iGPU | Intel CPU AMD dGPU |
|---|---|---|---|---|
| W-IJ | $47.72 \pm 3.85$ | $47.72 \pm 3.85$ | $47.72 \pm 3.85$ | $47.72 \pm 3.85$ |
| W-CLIJ | $47.39 \pm 3.64$ | $47.39 \pm 3.64$ | $47.74 \pm 3.89$ | $47.74 \pm 3.89$ |
| W-OPT-IJ | $46.19 \pm 3.9$ | $46.19 \pm 3.9$ | $46.19 \pm 3.9$ | $46.19 \pm 3.9$ |
| W-OPT-CLIJ | $46.64 \pm 3.62$ | $46.64 \pm 3.62$ | $47.01 \pm 3.87$ | $47.01 \pm 3.87$ |

Table 1.2: Mean $\pm$ standard deviation of measured signal intensities.

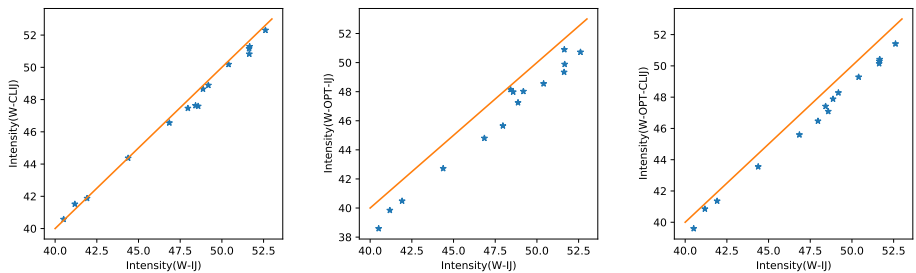

Figure 1.8: Scatter plots of measurements resulting from the original ImageJ macro workflow versus the CLIJ workflow (left), the optimized ImageJ workflow (center) versus the optimized CLIJ workflows (right). The orange line represents identity.

**Correlation analysis**

If two methods are supposed to measure the same parameter, they should produce quantitative measurements with high correlation on the same data set. To quantify this, Pearson's correlation coefficient R can be utilized. When applying this method to our data, R values were in all cases above 0.98 indicating high correlation. For visualising this, scatter plots are the method of choice, as shown in Fig. 1.8. Again, correlation is necessary, but not sufficient for proving method similarity.

**Equivalence testing**

For proving that two methods A and B result in equal measurements with given tolerance, statistical hypothesis testing should be used. A paired t-test shows, if observed differences are significant. Thus, a failed t-test is also necessary, but not sufficient to prove method similarity. A valid method for investigating method similarity is a combination of two one-sided paired t-tests (TOST). First, we define a lower and an upper limit of tolerable differences between method A and B, for example $\pm$ 5 %. Then, we apply one one-sided paired t-test to check if measurements of method B are less than 95 % compared to method A, and another one-sided t-test to check if measurements of method B are greater than 105 % compared to method A. Comparing the original workflow to the translated CLIJ workflow, the TOST showed that observed differences were within the tolerance (p-value < 1e-11).

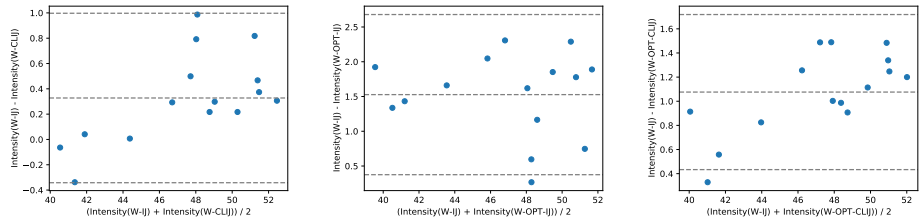

Figure 1.9: Bland-Altman plots of differences between measurements, resulting from the original ImageJ macro workflow versus the CLIJ workflow (left), the optimized ImageJ workflow (center), and the optimized CLIJ workflows (right). The dotted lines denote the mean difference (center) and the upper and lower bound of the 95% confidence interval.

**Bland-Altman analysis**

Another method of analysing differences between two methods is to determine a confidence interval, as suggested by Altman and Bland (1983). Furthermore, so-called Bland-Altman plots deliver a visual representation of differences between methods, as shown in Fig. 1.9. When comparing the original workflow to the CLIJ version, the mean difference was about 0.4, and differences between the methods lie within the 95% confidence interval [-0.4, 1]. The mean of the two methods ranges between 40 and 53. Thus, when processing our example dataset, the CLIJ workflow delivered intensity measurements of about 1% lower than the original workflow.

### 1.5.4 Benchmarking

After translating the workflow and assuring that the macro executes the right operations on our data, benchmarking is a common process to analyse the performance of algorithms.

**Fair performance comparison**

When investigating GPU-acceleration of image analysis procedures, it becomes crucial to obtain a realistic picture of the workflows performance. By measuring the processing time of individual operations on GPUs compared to ImageJ operations using CPUs, it was shown that GPUs typically perform faster than CPUs (Haase et al., 2020). However, pushing image data to the GPU memory and pulling results back take time. Thus, the transfer time needs to be included when benchmarking a workflow. The simplest way is to measure the time at the beginning of the workflow and at its end. Furthermore, it is recommended to exclude the needed time to load from hard drives, assuming that the hard-drive does not influence the processing time of CPUs or GPUs. After the *open()* image statement, the initial time measurement should be inserted:

```
start_time = getTime();
```

Before saving the results to disc, we measure the time again and calculate the time difference:

```
end_time = getTime();
print("Processing took " + (end_time-start_time) + " ms");
```

The *getTime()* method in ImageJ delivers the number of milliseconds since midnight of January 1, 1970 UTC. By subtracting two subsequent time measurements, we can calculate the passed time in milliseconds.

**Warm-up effects**

To receive reliable results, time measurements should be repeated several times. As shown in section 1.4, the first execution of a workflow is often slower than subsequent runs. The reason is the so-called warm-up effect, related to just-in-time (JIT) compilation of Java and OpenCL code. This compilation takes time. To show the variability of measured processing times between the original workflow and the CLIJ translation, we executed the workflows in loops for 100 times each. To eliminate resulting effects of different and subsequently executed workflows, we restarted Fiji after each 100 executions. From resulting time measurements, we derived a statistical summary in form of the median speedup factor. Visualized by box plots, we have an overview of the performance of the four different workflows, executed on four tested systems[3].

**Benchmarking results and discussion**

The resulting overview of processing time is given in Fig. 1.10. Dependent on the tested system, the CLIJ workflow results in median speedup factors between 1.5 and 2.7. These results must be interpreted with care. As shown earlier (Haase et al., 2020), workflow performance depends on many aspects, such as the number of operations and parameters, used hardware, and image size. When working on small images, which fit into the so-called Level-1 and Level-2 cache of internal CPU memory, CPUs typically outperform GPUs. Some operations perform faster on GPUs, such as applying convolution or other filters, which take neighboring pixels into account. By nature, there are operations which are hard to compute on GPUs. Such an example is the connected components labelling. As already described in section 1.5.2, we identified this operation as a bottleneck in our workflow. Afterwards, the optimized CLIJ workflow performed up to 5.5 times faster than the original. Hence, a careful workflow design is key to high performance. Identifying slow parts of the workflow and replacing them by alternative operations becomes routine when processing time is taken into account.

---

[3]https://github.com/haesleinhuepf/Preprint_ImageJ_Macro_CLIJ/blob/master/code/performance_comparison.ipynb

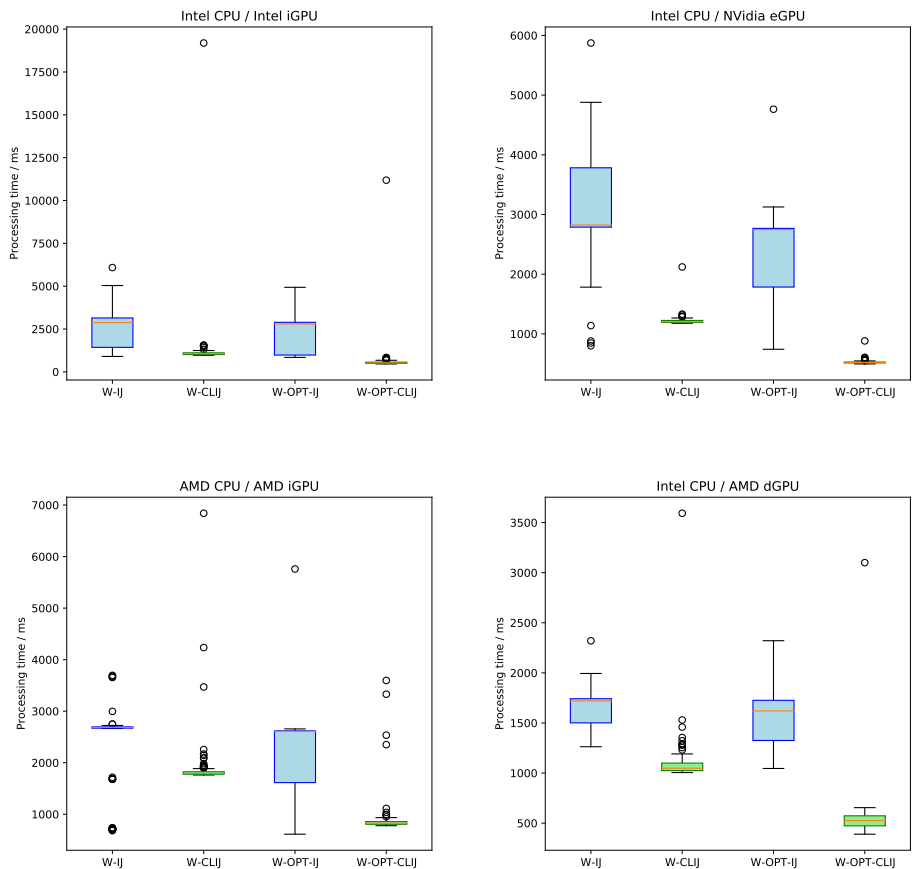

Figure 1.10: Box plots showing processing times of four different macros, tested on four computers. In case of the of classic ImageJ macro, blue boxes range from the 25th to the 75th percentile of processing time. Analogously, green boxes represent processing times of the CLIJ macro. The orange line denotes the median processing time. Circles denote outliers. In case of the CLIJ workflow, outliers typically occur while the first iteration, where compilation time causes the warm-up effect.

**Exercise 3**

Use the methods introduced in this section to benchmark the script presented in section 1.4. Compare the performance of the *mean* filter in ImageJ with its CLIJ counterpart. Determine the median processing time of both filters, including push and pull commands when using CLIJ.

## 1.6 Summary

The method of live-imaging, in particular recording long-term time lapses with large spatial resolution, is of increasing importance to study dynamic biological processes. Due to increased processing time of such data, image processing is the major bottleneck. In this chapter, we introduced one potential solution for faster processing, namely by GPU-accelerated image processing using CLIJ. We also demonstrated a step-by-step translation of a classic ImageJ Macro workflow. Obviously, GPU-acceleration is suited for particular use cases. Typical cases are

- processing of data larger than 10 MB per time point and channel,

- application of 3D image processing filters, such as convolution, mean, minimum, maximum, Gaussian blur,

- acceleration of workflows which take significant amount of time, especially, if processing is 10 times longer than loading and saving images,

- extensive workflows with multiple operations, consecutively executed on the GPU,

- and last but not least, utilizing sophisticated GPU-hardware with a high memory bandwidth, typically using GDDR6 memory.

When these conditions are met, speedup factors of one or two orders of magnitude are feasible. Furthermore, the warm-up effect is crucial. For example, if the first execution of a workflow takes ten times longer than subsequent executions, it becomes obvious that at least 11 images have to be processed to overcome the effect and to actually save time. When translating a classic workflow to CLIJ, some refactoring is necessary to follow the concept of processing granular units of image data by granular operations. This also improves readability of workflows, because operations on images are stated explicitly and in order of execution. Additionally, the shown methods for benchmarking and quality assurance can also be used in different scenarios as they are general method comparison strategies. GPU-accelerated image processing opens the door for more sophisticated image analysis in real-time. If days of processing time can be saved, it is worth to invest hours for learning CLIJ.

## 1.7 Solutions to the Exercises

**Exercise 1**

While applying image processing methods, the ImageJ Macro recorder records corresponding commands. This is an intuitive way to learn ImageJ Macro programming and CLIJ. After executing this exercise, the recorder should contain code like this:

```
open("/path/to/images/NPCsingleNucleus.tif");
selectWindow("NPCsingleNucleus.tif");
run("CLIJ2 Macro Extensions", "cl_device=[Intel(R) HD
   Graphics 630]");

// threshold otsu
image1 = "NPCsingleNucleus.tif";
Ext.CLIJ2_push(image1);
image2 = "threshold_otsu-936068520";
Ext.CLIJ2_thresholdOtsu(image1, image2);
Ext.CLIJ2_pull(image2);
```

It opens the data set, initializes the GPU, pushes the image to GPU memory, thresholds the image and pulls the resulting image back to show it on screen.

The Fiji search bar allows to select CLIJ methods. The corresponding dialog gives access to the CLIJ website, where the user can read about typical predecesssor and successor operations. For example, as shown in section 1.5.1 in Fig. 1.6, operations such as Gaussian blur, mean filter and Difference-Of-Gaussian are listed, which allow an improved segmentation, because they reduce noise.

**Exercise 2**

The recorded macro, adapted to print time traces, looks like this:

```
run("T1 Head (2.4M, 16-bits)");
run("CLIJ2 Macro Extensions", "cl_device=[Intel(R) UHD
   Graphics 620]");

// top hat
image1 = "t1-head.tif";
Ext.CLIJ2_push(image1);
image2 = "top_hat-427502308";
radius_x = 10.0;
radius_y = 10.0;
radius_z = 10.0;

// study time tracing of the Top Hat filter
Ext.CLIJ2_startTimeTracing();
Ext.CLIJ2_topHatBox(image1, image2, radius_x, radius_y,
   radius_z);
Ext.CLIJ2_stopTimeTracing();

Ext.CLIJ2_pull(image2);

// determine and print time traces
Ext.CLIJ2_getTimeTracing(time_traces);
print(time_traces);
```

The traced times, while executing the *Top Hat* filter on the T1-Head data set, are shown in Fig. 1.11. The *Top Hat* filter is a *minimum* filter applied to the original image, which is followed by a *maximum* filter. The result of these two operations is subtracted from the original. The two filters take about 60 ms each on the 16 MB large input image, the subtraction takes 5 ms. The *Top Hat* filter altogether takes 129 ms. *Top hat* is a technique to subtract background intensity from an image.

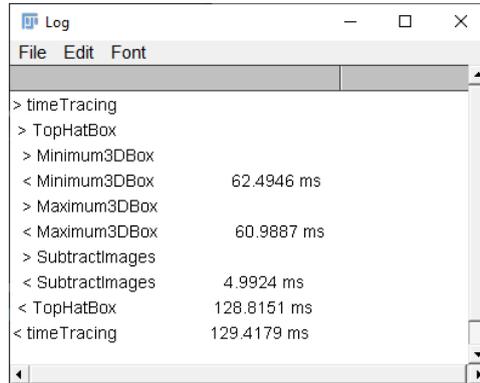

Figure 1.11: While executing the *Top Hat* filter, activated time tracing reveals that this operation consists of three subsequently applied operations: a *minimum* filter, a *maximum* filter and image subtraction.

**Exercise 3**

For benchmarking the mean 3D filter in ImageJ and CLIJ two example macros are provided online[4]. We executed them on our test computers and determined median execution times between 1445 and 5485 ms for the ImageJ filter and from 81 to 159 ms for the CLIJ filter, respectively.

## 1.8 Acknowledgments

We thank Gene Myers (CSBD / MPI-CBG) for constant support and giving us the academic freedom to advance GPU-accelerated image processing in Fiji. We also would like to thank our colleagues who supported us in making CLIJ and CLIJ2 possible in first place, namely Alexandr Dibrov (CSBD / MPI-CBG), Deborah Schmidt (CSBD / MPI-CBG), Florian Jug (CSBD/ MPI-CBG, HT Milano), Loïc A. Royer (CZ Biohub), Matthias Arzt (CSBD / MPI-CBG), Martin Weigert (EPFL Lausanne), Nicola Maghelli (MPI-CBG), Pavel Tomancak (MPI-CBG), Peter Steinbach (HZDR Dresden), and Uwe Schmidt (CSBD / MPI-CBG). Furthermore, development of CLIJ is a community effort. We would like to thank the NEUBIAS Academy [5]

---

[4] https://github.com/haesleinhuepf/Preprint_ImageJ_Macro_CLIJ/tree/master/code/exercise_3

[5] https://neubiasacademy.org/

and the Image Science community [6] for constant support and feedback.

Last but not least, we thank the German Federal Ministry of Research and Education (BMBF) for the support of R.H., under the code 031L0044 (Sysbio II).

## 1.9 Further Readings

On top of the given references in the main text, readers interested in state-of-the-art benchmarking approaches in high performance computing are recommended to read the overview given by Hoefler and Belli (2015). Furthermore, a research software engineers perspective on developing GPU-accelerated applications is also worth taking a closer look (van Werkhoven et al., 2020).

---

[6] https://image.sc/